\documentclass{appolb}
\usepackage{epsfig}
\def\be{\begin{eqnarray}}
\def\ee{\end{eqnarray}}

\begin{document}
\title{Random Matrix Filtering in Portfolio Optimization%
\thanks{Presented at Application of Random Matrices to Economy and Other
  Complex Systems, May 25-28, 2005, Krak\'ow, Poland}%
}
\author{G. Papp$^1$, Sz. Pafka$^{1,2}$, M.A. Nowak$^3$ and I. Kondor$^{1,4}$
\address{%
$^1$Institute of Physics, E\"otv\"os University, P\'azm\'any P. s. 1/a, H-1117
  Budapest, Hungary,\\
$^2$CIB Bank, Medve u 4-14, H-1027, Budapest, Hungary,\\
$^3$Institute for Physics, Jagellonian University, ul. Reymonta 4, 30-059
  Cracow, Poland,\\
$^4$Collegium Budapest - Institute for Advanced Study, Szenth\'aroms\'ag
u. 2, H-1014 Budapest, Hungary
}
}
\maketitle
\begin{abstract}
We study empirical covariance matrices in finance. Due to the limited amount of
available input information, these objects incorporate a huge amount of noise,
so their naive use in optimization procedures, such as portfolio selection,
may be misleading. In this paper we investigate a recently introduced
filtering procedure, and demonstrate the applicability of this method in a
controlled, simulation environment. 
\end{abstract}
  
\section{Introduction}
Investment decisions are governed by weighing risk vs. reward, that is
possible loss against expected return. Markowitz' classical portfolio
theory~\cite{Mar} assumes that the underlying stochastic process is 
multivariate normal with known returns and covariances. In practice, 
these parameters have to be determined from observations on the market.
Since the number of observations is necessarily limited, empirically determined 
parameters will always contain a certain measurement error. Even if we
disregard the notoriously hard problem of estimating returns and concentrate
solely on the covariances, we still run into a problem of serious information
imbalance: the size $N$ of typical bank portfolios is too large compared to
the amount of information contained in the finite-length time series available
for the assets in the portfolio. As the number of input data is $N\times T$,
where $T$ is the length of the time series, whereas the number of data needed
for the construction of the covariance matrix is ${\cal O}(N^2)$, we expect that the
quality of the estimate depends essentially on the ratio $N/T$ and that the
error goes to zero only in the limit of very small $N/T$.  Now the problem is
that $N/T$ is never sufficiently small in practice, in fact, it may easily
become larger than unity, the threshold value where the covariance matrix
becomes singular and the portfolio selection problem meaningless. 

Over the past decades a large number of different techniques have been
developed to tackle this problem and reduce the effective dimension of large
bank portfolios~\cite{Gruber}. Our purpose here is to apply a recently
introduced filtering procedure~\cite{Clean} in a well controlled simulation
setting where the efficiency of the method can be reliably tested.

In order to determine the optimal portfolio, one has to invert the covariance
matrix. Since this has, as a rule, a number of small eigenvalues, any
measurement error will get amplified and the resulting portfolio will be
sensitive to the noise. In order to study the effect of noise, we start from a
known correlation matrix, dress it with noise, and reestablish the results
of~\cite{PK1}. Next, we apply the cleaning procedure of~\cite{Clean} to the
empirical covariance matrix, and investigate the improvement of the result
compared with the original, unfiltered theory.

\section{Results and discussion}
\subsection{Model correlation matrix}
Let us start from a known covariance matrix, ${\bf C}$, of size $N\times
N$, representing the true correlation between $N$ instruments making up the
portfolio. The portfolio weights $w_i$ , ($i=1,..,N$), satisfy the constraint
$\sum_i w_i = 1$, and we assume that short-selling is allowed, i.e. some of 
the weights can be negative. For the sake of simplicity, we do not impose any 
further conditions on the weights (like e.g. the usual constraint on expected
returns, which cannot be determined on a daily horizon with any reliability
anyhow), and concentrate on the minimal risk portfolio. In a Gaussian world
the natural measure of risk is the portfolio variance which is then our
objective function to be minimized,
\be
  R^2 = \sum_{i,j=1}^N\ w_i\, {\bf C}_{ij}\, w_j\,.
\ee
After some trivial algebra one finds the optimal weights as
\be
  w_i^{*} = \frac{\sum_{j=1}^N\, {\bf C}^{-1}_{ij}}%
    {\sum_{i,j=1}^N\, {\bf C}^{-1}_{ij}} \,.
\ee

\begin{figure}[ht]
\centerline{
 \includegraphics[width=0.4\textwidth]{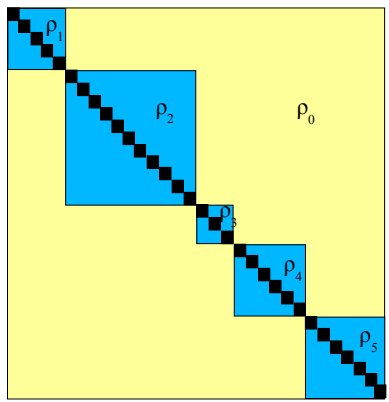}
}
\caption{Structure of the market-plus-sectors model correlation matrix. Correlation with the market is $\varrho_0$, while the correlation inside sector $i$ is $\varrho_i$ ($i=1,2,..$).}
\label{fig-sector}
\end{figure}

It is natural to assume that stocks of companies belonging to a given
industrial sector are more strongly correlated than those belonging to
different sectors. Accordingly, we expect that the covariance matrix displays
a block diagonal structure. For simplicity, we assume that the elements
outside the diagonal blocks (that describe some general correlation with the
whole market) are all equal and non-negative, $\varrho_0 \ge 0$, and the
elements $\varrho_i$, ($i=1,2,..$), describing intra-sector correlations in
the diagonal blocks are constants within each block, and larger than those
outside the blocks,  $\varrho_i\ge \varrho_0$. The model just described is the
same as the one introduced by Noh~\cite{Noh}. For the sake of simplicity
again, we study a case when the correlation and covariance matrices are the
same, i.e. we set the variance of individual instruments to unity. The
structure of the correlation matrix, which we will refer to as the
market-plus-sectors-model in the following, is then given by the pattern shown in Figure~\ref{fig-sector}.
Such a matrix, containing $M$ sectors, possesses $M$ small eigenvalues given by 
$1-\varrho_i < 1$, $i=1,2,..$. The corresponding eigenvectors will be strongly
localized, having only two nonzero elements (of equal absolute value but
opposite sign). Their multiplicity is $N_i-1$, $i=1,2,..$ (where $N_i$ is the
number of entries within sector $i$), i. e. the total multiplicity of the
small eigenvalues is $N-M$. In addition, there are $M$ large eigenvalues
($\lambda> 1$), typically singlets, that depend on all the parameters of the
model: $\varrho_0$, $\varrho_i$, and $N_i$. That is, an $M$ sector matrix has
$2M$ different eigenvalues. By virtue of the Frobenius-Perron theorem, the
largest eigenvalue will necessarily be a singlet of ${\cal O}(N)$, with an
eigenvector having all positive components. This mode can then be identified
with the whole market. 

On the whole, this simple model reproduces all the main features observed in
the spectra of real-life empirical covariance matrices~\cite{Bouchaud}. It
will also be useful to consider two special cases of the model: When all the
$\varrho_i$'s are the same and equal to $\varrho_0$, the block structure
disappears, and we are left with a one-factor model, where only covariances
with the market matter. The spectrum of the corresponding covariance matrix
consists of merely two values, a large eigenvalue of order $N$, and an
$(N-1)$-fold degenerate small eigenvalue, $1-\varrho_0 < 1$.
As a further simplification, we can even drop $\varrho_0$, and end up with the
unit matrix as the most trivial covariance matrix conceivable.

\subsection{Empirical correlation}
Assuming that we have chosen one of the above models, the market-plus-sectors model, the one-factor model, or just the unit matrix, we can construct the corresponding empirical correlation matrix from them as follows: we
generate finite time series from the true correlation matrix
${\bf C}$, 
\be
  x_{it} = \sum_j\, {\bf A}_{i,j}\ y_{jt} \quad t=1,..,T \,,
\ee
where ${\bf A}$ is the Cholesky decomposition of the true correlation matrix 
${\bf C} = {\bf A} {\bf A}^T$, and $y_{jt}$ is a random Gaussian variable
with mean zero and variance 1 at time $t$.

Then the empirical correlation matrix is given by the usual estimator as
\be
  {\bf C}^{(e)} = \frac1T\, \sum_{t=1}^T\ x_{it} x_{jt} \,.
\ee

The resulting empirical covariances will fluctuate from sample to sample. 
The main effect of this noise will be to resolve the degeneracy of the small
eigenvalues, so that for a large enough matrix they form a quasi-continuous
band. For $N$ and $T$ going to infinity so that $r=N/T$ is fixed and smaller
than one, the spectral density of the small eigenvalues will be given by the
Marchenko-Pastur spectrum~\cite{Marchenko}. (For $r$ larger than one, an
additional Dirac-delta appears at the origin.) For small enough $r$'s the
large
eigenvalues remain relatively unaffected by the noise, but as $r$ grows and
approaches unity, the effect of noise becomes dramatic, as we demonstrate
below.

The Markowitz-weights corresponding to the empirical covariance matrix are
\be
  w_i^{(e)} = \frac{\sum_{j=1}^N\, {{\bf C}^{(e)}}^{-1}_{ij}}%
    {\sum_{i,j=1}^N\, {{\bf C}^{(e)}}^{-1}_{ij}} \,.
\ee

Now we can evaluate the risk associated with this choice of the
portfolio. A possible way to characterize the effect of measurement error 
is to evaluate the variance by using the true correlation matrix ${\bf
  C}$ with the weights calculated from the empirical one, ${\bf C}^{(e)}$,
\be
  {R^{(e)}}^2 =  \sum_{i,j=1}^N\ w_i^{(e)}\, {\bf C}_{ij}\, w_j^{(e)}.
\ee
Since the empirical weights are not optimal, we always have
${R^{(e)}}^2 \ge R^2$.

In the following we will use
\be
  q_0^2 = \frac{R^{(e)}}{R} \quad \ge 1
\ee
as a measure of the effect of noise on portfolio selection.
\begin{figure}[ht]
\centerline{
  \rotatebox{-90}{
   \includegraphics[width=0.45\textwidth]{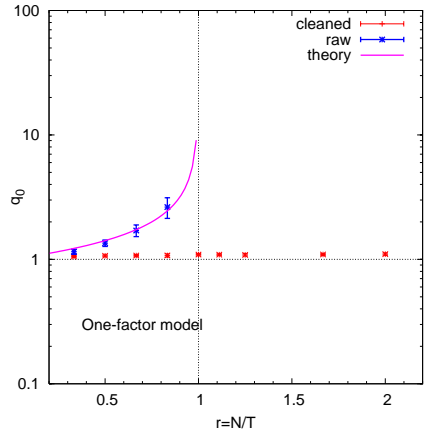}
  }
}
\caption{One factor model results for $N=100$, $q_0$ vs. $r$: optimization
  with empirical 
  correlation matrices (errorbars), random matrix result (solid line), and
  optimization with the cleaned one-sector correlation matrix (stars). For
  $r>1$ the standard Markowitz theory is not applicable.}
\label{fig-q0}
\end{figure}

$q_0$ can be easily evaluated for the special case when the true covariance
matrix is just the unit matrix. Then the empirical covariance matrix will be a
random matrix with a spectral density fast converging to the Marchenko-Pastur
spectrum~\cite{Marchenko}:
\be
  \varrho(\lambda) = \frac1{2r\pi\lambda}\,
  \sqrt{(\lambda-\lambda_<)(\lambda_>-\lambda)}
\label{eq:wishart}
\ee
where $r=N/T$, $\lambda_{<,>}=(1\pm\sqrt{r})^2$.  
Evaluating $q_0$ in the diagonal representation, we get
\be
  q_0^2 = \frac{\int\!d\lambda\ \frac1{\lambda^2}\varrho(\lambda)}%
   {\left(\int\!d\lambda\ \frac1{\lambda}\varrho(\lambda)\right)^2} 
   = \frac1{1-r} \,.
\ee

This strikingly simple result, dating back to a discussion between the present
authors, was first published in~\cite{PK1}. It remains valid up to 
${\cal O}(1/N)$ corrections also for the one-factor model, and, within corrections controlled by the size of the sectors, also for the market-plus-sectors model. It tells us that as the size $N$ of the portfolio grows and approaches the length of the time series $T$, the error in the portfolio diverges. While it is a commonplace that at the threshold $N=T$ the portfolio problem becomes meaningless (the covariance matrix looses its positive definite character), it does not seem to have been noticed in the quantitative finance literature earlier that the error can be given by such a simple exact formula.

A comparison between the theoretical prediction and the simulation is displayed
in Figure~\ref{fig-q0}: the agreement is perfect. Concerning the relevance of our simple result for real markets, one has to realize that it has been derived on the basis of idealized conditions: perfect stationarity of the process and Gaussian distribution of returns. Neither of these holds true on real markets, therefore we believe that our formula is a lower bound for the error in real-life portfolios.

Given the fact that $r$ is never small in practice, and, in fact, it may even go beyond the critical value $r=1$, it is imperative that some sort of filtering or cleaning procedure be applied, in order to reduce the effect of noise. A number of these techniques is available in the literature~\cite{Gruber}. Each of them corresponds to injecting some external information, additional to the time series data, into the empirical covariance matrix. The procedure proposed recently in~\cite{Clean} requires that we make an educated guess concerning the structure of the market. We are going to test its performance in the next section.

\subsection{Cleaned correlation}
Studies of real financial empirical matrices~\cite{Bouchaud} have revealed
that they only have a relatively small number of large eigenvalues (in the
case of the S\&P500 less than 20), the rest are small, and conform rather well to
the Marchenko-Pastur pattern~(\ref{eq:wishart}). This must mean that the
number of relevant sectors is fairly small compared with the size of the
portfolio. Our task is therefore to reconstruct the true correlation matrix
assuming a structure with a few sectors.

\begin{figure}[ht]
\centerline{
  \rotatebox{-90}{
   \includegraphics[width=0.45\textwidth]{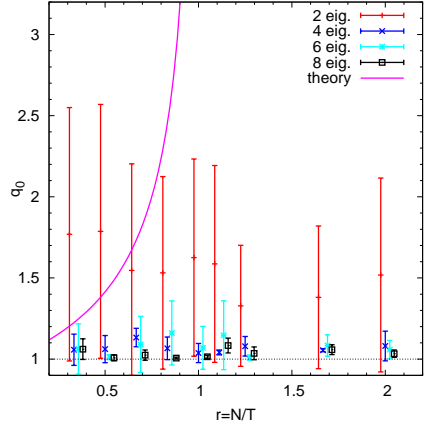}
  }
}
\caption{Market-plus-sectors model results for $N=100$, $q_0$ vs. $r$, for
  different numbers of eigenvalues reconstructed. Above 4 eigenvalues the
  result does not change anymore. }
\label{fig-two}
\end{figure}

The general theory of the cleaning procedure of the empirical correlation
 matrix dressed with Gaussian noise is described in~\cite{Clean}. Let $G(Z)$
 be the resolvent for the cleaned correlation matrix
 \be
  G(Z) = \frac1N \, {\mathrm Tr} \frac1{Z-{\bf C}} \ ,
\ee
with a similar formula for the the resolvent $g(z)$ of the empirical correlation matrix. Then the relation between the two is expressed as
\be
  z\, g(z) = Z\, G(Z), \quad \mbox{with} \quad z = \frac{Z}{1-r+r Z G(Z)} \,.
\ee
This can be translated into a relation between the corresponding moments, and
from the knowledge of $2 M-1$ moments one is able to reconstruct $M$ sectors
 for the true correlation matrix. We also note that this procedure, applied in
 the reverse direction, allows one to calculate the spectral density for more
 complicated scenarios, for correlated random matrices. As a result, the
 eigenvalue spectrum will slightly, but noticeably change, and come to a
 closer agreement with the one observed in finance. 
The procedure described in~\cite{Clean} allows one to reconstruct the
eigenvalues only, but not the eigenvectors. Our aim here is to reconstruct
the true correlation matrix using the cleaned eigenvalues and the empirical
eigenvectors. The question is whether such a procedure can lead to any improvement?

We present our result in Figures~\ref{fig-q0},\ref{fig-two}, for the
one-factor model and the market-plus-sectors model, respectively. The true correlation matrix is well reconstructed in both cases, the portfolio built from the (cleaned) empirical data is suboptimal by only 5-10\%. Thus, the cleaning
procedure leads to a very substantial improvement compared with the naive use
of the empirical covariance matrix, and allows the optimization to be
performed even in the range $N>T$. 

The cleaning procedure does not determine the number of sectors, or of the
eigenvalues to be searched for, it is a parameter of the fitting. We tested
the method with different numbers of eigenvalues (i.e. different numbers of
sectors), and observed the changes in performance depending on the trial
structure. We find that there is a clear saturation after one reaches the
number of eigenvalues corresponding the number of sectors in the true
correlation matrix, and a further increase of the number of searched
eigenvalues does not changes the result. This allows one to determine the
proper number of sectors by looking for the signature of saturation.

On the other hand, no problem arises if one sets the number of searched
eigenvalues higher than required: the cleaning procedure will return less
independent eigenvalues, saturating at the maximal number allowed by the
number of sectors.

In conclusion, we have performed a preliminary study of the effect of the
random matrix based cleaning described in~\cite{Clean} on the optimization of
financial portfolios. We have found that the method works very efficiently in
an artificial test environment, reproducing nearly perfectly a one-factor
model scenario, and leading to a huge improvement in a market-plus-sectors
model with a moderate number of sectors. It is obvious that before a final
judgment can be passed on the method, a more detailed study of its various
aspects should be performed, extending also to a possible cleaning of the
eigenvectors. Furthermore, a careful comparison of the efficiency of the
method with other filtering procedures proposed in the literature is clearly
necessary.

\section*{Acknowledgements}
Two us (I.K and Sz.P.) are grateful for the hospitality of the Institute for
Theoretical Sciences - A Joint Institute of Argonne National Laboratory and
University of Notre Dame, where part of this manuscript has been prepared.
G.P. acknowledges the support of Hungarian OTKA grant T047050.

\end{document}